\documentclass[%
reprint,
superscriptaddress,
groupedaddress,
showkeys,
nofootinbib,
nobibnotes,
bibnotes,
amsmath,amssymb,
aps,
]{revtex4-1}

\usepackage{graphicx}% Include figure files
\usepackage{dcolumn}% Align table columns on decimal point
\usepackage{bm}% bold math
\usepackage{hyperref}% add hypertext capabilities
\usepackage{graphicx,slashed,hyperref,color,subfig}
\usepackage{amssymb}
\usepackage[normalem]{ulem}
\usepackage[utf8]{inputenc}
\hypersetup{
   bookmarks=true,         % show bookmarks bar?
   unicode=true,          % non-Latin characters in Acrobat�s bookmarks
   pdftoolbar=true,        % show Acrobat�s toolbar?
   pdfmenubar=true,        % show Acrobat�s menu?
   pdffitwindow=false,     % window fit to page when opened
   pdfstartview={FitH},    % fits the width of the page to the window
   pdftitle={My title},    % title
   pdfauthor={Author},     % author
   pdfsubject={Subject},   % subject of the document
   pdfcreator={Creator},   % creator of the document
   pdfproducer={Producer}, % producer of the document
   pdfkeywords={keyword1} {key2} {key3}, % list of keywords
   pdfnewwindow=true,      % links in new PDF window
   colorlinks=true,       % false: boxed links; true: colored links
   linkcolor=blue,          % color of internal links (change box color with linkbordercolor)
   citecolor=magenta,        % color of links to bibliography
   filecolor=magenta,      % color of file links
   urlcolor=cyan           % color of external links
}

\def\lsim{\mathrel{\rlap{\lower4pt\hbox{\hskip1pt$\sim$}}
    \raise1pt\hbox{$<$}}}         %less than or approx. symbol
\def\gsim{\mathrel{\rlap{\lower4pt\hbox{\hskip1pt$\sim$}}
    \raise1pt\hbox{$>$}}}         %greater than or approx. symbol

\begin{document}

%\preprint{APS/123-QED}

\title{Higgs Inflation and the Refined dS Conjecture}% Force line breaks with \\
%\thanks{This work was supported in part by the National Research Foundation of Korea (NRF) grant funded by the Korean government (MSIP) (No. 2016R1A2B2016112) and (NRF-2018R1A4A1025334)
%}%

\author{Dhong Yeon Cheong}
%\email{dhongyeon@yonsei.ac.kr}
%\affiliation{Department of Physics \& IPAP, Yonsei University, Seoul 03722 Korea}

\author{Sung Mook Lee}
%\email{sungmook.lee@yonsei.ac.kr}
%\affiliation{Department of Physics \& IPAP, Yonsei University, Seoul 03722 Korea}

\author{Seong Chan Park}
\email{sc.park@yonsei.ac.kr}
\affiliation{Department of Physics \& IPAP \& Lab for Dark Universe, Yonsei University, Seoul 03722 Korea}

%% \altaffiliation[Also at ]{Physics Department, XYZ University.}%Lines break automatically or can be forced with \\

\date{\today}% It is always \today, today,

\begin{abstract}
 The refined de Sitter derivative conjecture provides constraints to potentials that are low energy effective theories of quantum gravity. It can give direct bounds on inflationary scenarios and determine whether the theory is in the Landscape or the Swampland. We consider the `Higgs inflation' scenario taking the refined de Sitter derivative conjecture into account. Obtaining the critical lines for the potential, we find a conjecture parameter space in which the `Higgs inflation' is to be in the Landscape. Comparing with the model independent observational bounds from recent data we find that the observational bounds represent the Higgs inflation can be in the Landscape.
\end{abstract}

\pacs{Valid PACS appear here}% PACS, the Physics and Astronomy
%Use showkeys class option if keyword
\maketitle

%%%%%%%%%%%%%%%%%%%%%%%%%%%%%%%%%%%%%%
\section{ Introduction}
\label{sec:introduction}
%%%%%%%%%%%%%%%%%%%%%%%%%%%%%%%%%%%%%%

Recent papers proposed new Swampland conjecture criteria regarding the derivative of the scalar field potentials. The so called `de Sitter derivative conjecture' states that the total scalar potential $V\left(\phi\right)$ of a low effective theory consistent to a reasonable quantum gravity theory needs to satisfy at least one of the following conditions \cite{Garg:2018reu, Obied:2018sgi, Ooguri:2018wrx}.  
\begin{eqnarray}
&M_P ||\nabla V|| \geq c_{1}{V}, \\ 
&M_P^2 \text{min}\left(\nabla_i \nabla_j V\right) \leq -c_{2} V,
\end{eqnarray}
where $c_{1}$, $c_2$ are ${\mathcal{O}}(1)$ positive constants.
\footnote{Taking $V'\sim \Delta V/\Delta \phi$, we get $\Delta \phi/M_P \lsim 1/c_1$ for $\Delta V/V <1$ from the first condition. We observe that this result is essentially compatible with  the Lyth bound~\cite{Copeland:1994vg}.}

The first conjecture, which corresponds to the initial ``dS derivative conjecture" had severe tension with numerous phenomenological models, both in particle physics and inflation. In particular, the single derivative conjecture ruled out the de Sitter (dS) vacua of a consistent theory as $||\nabla V||=0$ but $V>0$ for a dS vacuum, even though the dS vacuum is the vacuum energy solution of Einstein's equation with a positive cosmological constant. Hence the cosmological constant scenario is excluded when taking this conjecture to be true. 
In order to explain the currently observed universe, one may alternatively adopt the quintessence field with an exponentially decaying potential $V_{Q} \left(Q\right) = \Lambda_{Q}^{4} e^{-c_{Q} Q}$ %
which allows the dynamical vacuum energy with the present values of the quintessence field and the scale parameter as 
$Q \sim 0$ and $\Lambda_Q \sim \mathcal{O}(1)$ meV.  Interestingly, the swampland conjecture then restricts the range of the unknown parameter $c_Q$ as $||\nabla V_Q||  = c_Q V_Q \geq  c_{*} V_Q$  or $c_Q \geq c_*$.

The electroweak sector and inflation needed to be modified when ``only" taking the first derivative conjecture into account. The electroweak sector with the Higgs potential  $V_{H} = {\lambda}\left(h^2 - v^2\right)^2$ violated the conjecture at the local maximum~\cite{Denef:2018etk, Murayama:2018lie, Han:2018yrk}, and conventional single field slow-roll inflation also contained severe tension with observational parameters, and needed modifications to satisfy the conditions \cite{Agrawal:2018own, Achucarro:2018vey, Kehagias:2018uem, Kinney:2018nny}.  We also see many papers considering various constructions and implications of the conjecture~\cite{
	Andriot:2018wzk,
	Andriot:2018ept,
	Ben-Dayan:2018mhe,
	Heisenberg:2018yae,
	Akrami:2018ylq, 
	Kinney:2018nny,
	Brahma:2018hrd,
	Brandenberger:2018wbg,
	DAmico:2018mnx,
	Visinelli:2018utg, 
	Moritz:2018ani, 
	Bena:2018fqc,
	Brandenberger:2018xnf,
	Halverson:2018cio, 
	Ellis:2018xdr,
	Anguelova:2018vyr,  
	Lin:2018kjm,
	Hamaguchi:2018vtv,
	Kawasaki:2018daf, 
	Motaharfar:2018zyb,
	Dimopoulos:2018upl,
	Odintsov:2018zai,
	Ashoorioon:2018sqb, 
	Das:2018rpg,
	Antoniadis:2018ngr,
	Wang:2018kly,
	Agrawal:2018rcg,
	Heckman:2018mxl}.

The addition of the second conjecture weakened the condition when the first conjecture was the sole introduction. This conjecture can be easily satisfied through generic potentials where the field value is in low scale $\Delta\phi \ll M_P$.\footnote{This condition is consistent within other swampland conjectures, specifically the field range conjecture \cite{Ooguri:2006in}.}
\begin{align}
M_P^2 \frac{\nabla_i \nabla_j V}{V} \sim -\frac{M_P^2}{\Delta \phi^2} \ll -c_2 \sim -\mathcal{O}\left(1\right).
\end{align}
The original EW sector, axions, and other phenomenological situations with high tension with the first condition are made plausible theories in the Landscape by the addition to the conjecture. 

Previous single field, slow roll inflation models are also revived by this introduction.  However, in contrast to the low energy regime, inflationary dynamics occur at high energy scales, which do not always satisfy the conjecture. Hence the conjecture does provide constraints to inflationary models by providing possible values for $c_{1}, c_{2}$~\cite{Fukuda:2018haz}. Recently, generic methods of analyzing a monotonic potential with an inflection point have been developed~\cite{Park:2018fuj} and minimal gauge inflation~\cite{Gong:2018jer} has been considered in detail~\footnote{In the finalizing stage of this paper, \cite{Chiang:2018lqx} has appeared stating model independent bounds on inflationary models, which have some overlaps with our results.}. 

We now turn our interest to the Higgs inflation scenario. The goal is to find the constraints for constants $c_{1}, c_{2}$ given from the Higgs inflation potential, and compare with model independent parameter spaces. In nature, we have already observed the Higgs field at the LHC~\footnote{Strictly speaking, the quanta of the field, $H$.}, and it is proposed to be the only scalar field in the Standard Model. The Higgs inflation interprets the Higgs as the inflaton\cite{Bezrukov:2007ep}, where the Higgs field has a nonminimal coupling to gravity. At the tree-level, the model predicts the spectral index of the primordial curvature power spectrum ($n_s\approx 0.965$) and tensor-to-scalar ratio ($r\approx 0.003$) in agreement with the latest observational data~\cite{Ade:2018gkx, Akrami:2018odb, Akrami:2018odb, Aghanim:2018eyx}. The effect of quantum corrections have been extensively studied in Refs.~\cite{Bezrukov:2009db, Hamada:2014iga, Hamada:2014wna,Bezrukov:2014ipa,Bezrukov:2014bra, Bezrukov:2017dyv}. The Higgs inflation scenario may also allow significant primordial black hole production (PBH) and explain the dark matter problem~\cite{Garcia-Bellido:2017mdw,Ezquiaga:2017fvi} (also see \cite{Bezrukov:2017dyv}). 

%\begin{figure}[tbp]
%\centering % \begin{center}/\end{center} takes some additional vertical space
%\includegraphics[width=.45\textwidth]{fig_higgs_inf_pot.png}
% "\includegraphics" is very powerful; the graphicx package is already loaded
%\caption{The potential diagram of the Higgs inflation($\lambda = 0.13$, $\xi = 17000$)}
%\label{fig:pot} 
%\end{figure}

%Thus we would examine this inflation scenario's possible $c_1 - c_2$ parameter space. 
This letter is composed as follows: in Sec.~\ref{sec:hi} we introduce our model set-up where the nonminimally coupled Higgs field and the Higgs potential are examined, and we calculate the parameters from the dS derivative conjecture from this specific potential.
In Sec.~\ref{sec:pheno}, we identify cases in which the Higgs potential satisfy the conjecture and obtain the parametric bounds of $c_1, c_2$. We then compute the model independent observational bounds and compare the overlapping regions with the Higgs potential. We conclude in Sec.~\ref{sec:conclusion}.

%=====================================================
\section{Nonminimally coupled Higgs Inflation and Swampland Parameters}
\label{sec:hi}
%=====================================================

In the unitary gauge, we write the ${\rm SU}(2)$ doublet Higgs as  $H = (0, v+h)^T/\sqrt{2}$ with the vacuum expectation value $v\approx 246$ GeV from the Fermi constant, $G_F =1/\sqrt{2} v^2 \approx 1.16 \times 10^{-5}$ GeV. 
The action of a general inflation (with the reduced Planck scale, $M_P=1$) can be written as

\begin{eqnarray}
S_{J}
&=& \int d^4x \sqrt{-g_J}\left[ \frac{1+2\xi H_J^\dagger H_J}{2}R_J - |D_\mu H_J|^2 -V_J(H_J)\right] \nonumber \\
&=&\int d^4x \sqrt{-g_J}\left[ \frac{1+\xi h_J^2}{2}R_J - \frac{1}{2}|D_\mu h_J|^2 -V_J(h_J)\right].
\label{eq:V}
\end{eqnarray}
where $D_\mu$ is the covariant derivative of the SM gauge interactions. 
We take the conformal transformation to shift the frame from the Jordan frame to the Einstein frame by introducing the metric $g_{\mu \nu} = \Omega^2 {g_{\mu \nu}^{J}}$ where the factor $\Omega^2 = 1+\xi h_{J}^2$.  Then the Ricci scalar transforms as 
$R_J = \Omega^2\left[ R + 3 \Box \ln \Omega^2 - \frac{3}{2} \left(\partial \ln \Omega^2 \right)^2\right]$. Hence the action becomes canonical in the Einstein frame with the form of 
%%%%%%%%%%%%%%%%%%%%
\begin{eqnarray}
S 
&=&\int d^4 x \sqrt{-g}\left[ \frac{1}{2}R -\frac{1}{2}F(h_J)|D_\mu h_J|^2 - \frac{V_J\left(h_J\right)}{\Omega^4}\right]\nonumber \\
&=&\int d^4 x \sqrt{-g}\left[\frac{1}{2}R -\frac{1}{2} |D_\mu h|^2  - V\right],
\end{eqnarray}
%%%%%%%%%%%%%%%%%%%%
where $F(h_J) =\frac{\Omega^2 + 6\xi^2 h_J^2}{\Omega^4} = \frac{1+\left(\xi+6\xi^2\right)h_J^2}{(1+\xi h_J^2)^2}$, $V = V_J/\Omega^4$ and  the canonically normalized field in the Einstein frame $h$ is obtained by solving the following:
\begin{eqnarray}
\frac{d h}{dh_J} = \sqrt{F(h_J)} = \sqrt{\frac{\Omega^2 + 6\xi^2 h_J^2}{\Omega^4}}
\label{eqn:htrans}
\end{eqnarray}
%%%%%%%%%%%%%%%%%%%%%%%%%
in the metric form.
The Higgs inflation potential in the Einstein frame becomes:
\begin{align}
V\left(h\right) = \frac{V_J}{\Omega^4} = \frac{\lambda \left(h_J(h) ^2- v^2\right)^2}{\left(1+\xi h_J(h)^2\right)^2}.
\label{fig:VJ}
\end{align}
Transforming frames according to Eq (\ref{eqn:htrans}), the potential in Einstein frame is approximately expressed as
\begin{eqnarray}
V\left(h\right) \approx \begin{cases}
\frac{\lambda}{\xi^2}\left( 1+e^{-\sqrt{\frac{2}{3}}h}\right)^{-2}  & \text{at}~~\sqrt{\xi} h_J(h)\gg 1, \label{eqn:vapprox}\\ 
{\lambda}\left(h^2-v^2\right)^2 & \text{at}~~\sqrt{\xi}h_J(h)\ll 1.
\end{cases}
\end{eqnarray}
\\
%
%The parameters $\xi$ and $\lambda$  determine the dynamics of inflation. \\

From this potential we define functions corresponding to the constants $c_1, ~ c_2$ from the dS swampland conjecture:
	\begin{eqnarray}
	F_1\left(h\right) 
	&\equiv& \frac{|dV/dh|}{V} ,\label{eqn:F1}\\
	F_2\left(h\right) 
	&\equiv& \frac{d^2V/dh^2}{V}.\label{eqn:F2}
	\end{eqnarray}
These functions evaluate the 1st and 2nd derivative conjecture until a desired field value
\begin{align}
F_1\left(h\right) \geq c_1 & \Leftrightarrow h \leq\phi^* = F_1^{-1} \left(c_1\right), 
\label{eq:cond1}\\
F_2\left(h\right) \leq -c_2 & \Leftrightarrow h \in [ \phi^\star_{\text{min}},\phi^\star_{\text{max}} ], 
\label{eq:cond2}
\end{align}
where $\phi_{\text{min}}^{\star}, \phi_{\text{max}}^\star$ are the allowed minimum and maximum field values for a certain $c_2$, and are determined by $F^{-1}_2(-c_2)$. 

\begin{figure}[tbp]
\centering % \begin{center}/\end{center} takes some additional vertical space
\includegraphics[width=.47\textwidth]{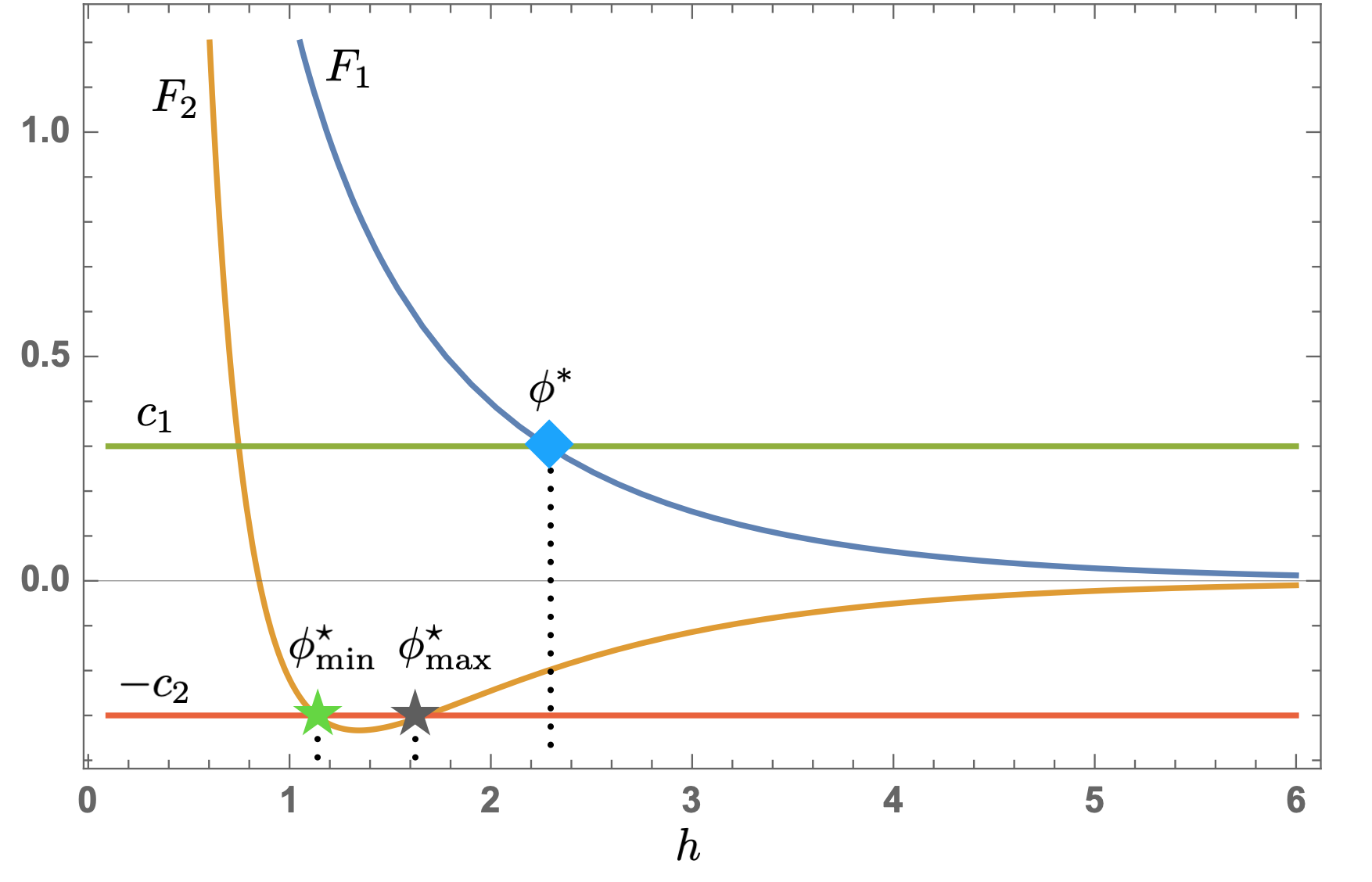}
% "\includegraphics" is very powerful; the graphicx package is already loaded
\caption{Functions $F_1\left(h\right),F_2\left(h\right)$ with parameters $\lambda = 0.13$, $\xi = 17000$.}
\label{fig:c1} 
\end{figure}
The functions are plotted in Fig~\ref{fig:c1}. Notice that the function $F_2$ is not a monotonic function, in contrast to the minimal gauge inflation case~\cite{Gong:2018jer,Park:2007sp} as well as the natural inflation case~\cite{Freese:1990rb}. This results in $\phi_{\text{min}}^{\star}, \phi_{\text{max}}^\star$ for one $c_{2}$ as expressed above. 
As the condition in Eq.~(\ref{eq:cond1}) provides the upper bound of $h$ at $F_1^{-1}(c_1)$ and the condition in Eq.~(\ref{eq:cond2}) provides a finite allowed region, $h \in [ \phi_{\text{min}}^{\star}, \phi_{\text{max}}^\star]$, there is no parametric region in $(c_1,c_2)$ which allows the whole field space, $h \in \left[0,\infty\right)$, satisfying the dS conjecture. Although this looks disappointing at first glance, what we actually need for a successful inflationary dynamics is not requesting the whole region of the field space but the space in which the inflationary dynamics takes place: $h \lsim h_*$ where we may choose $h_*=h^{N_e=60(50)}$  for a large enough number of e-folds, $N_e=60(50)$ assuming that the potential gets corrected by e.g. higher order operators, ${\cal O}(h^6/M_P^2)$~\cite{Hamada:2014iga, Hamada:2014wna}.
If we accept this phenomenological requirement, then we can still find a reasonable parameter space in $(c_1,c_2)$ for $h\lsim h_*$ as we will explicitly show in the next section.

\begin{figure*}[tbp]
\centering % \begin{center}/\end{center} takes some additional vertical space
\includegraphics[width=0.47\textwidth]{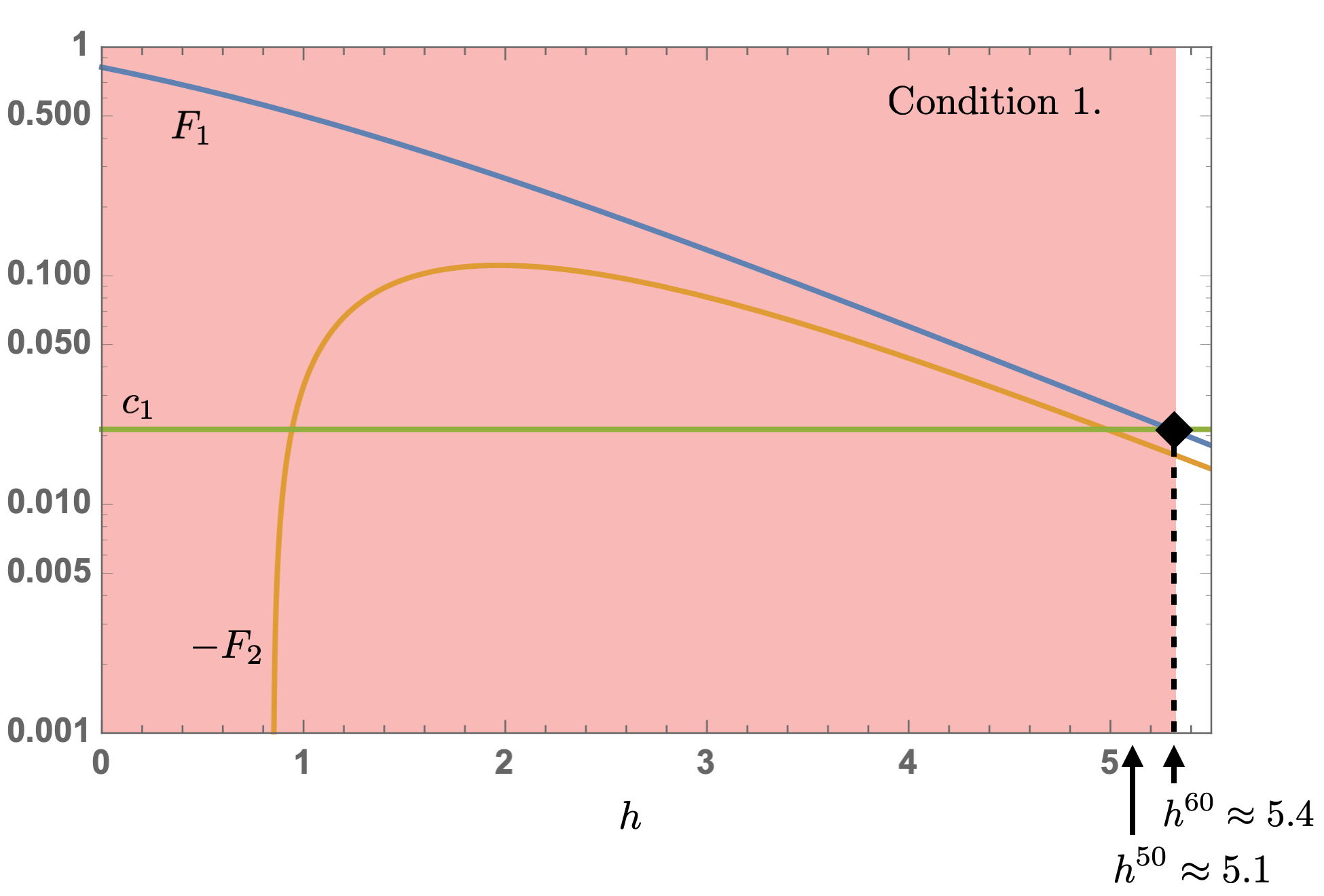}
\includegraphics[width=0.47\textwidth]{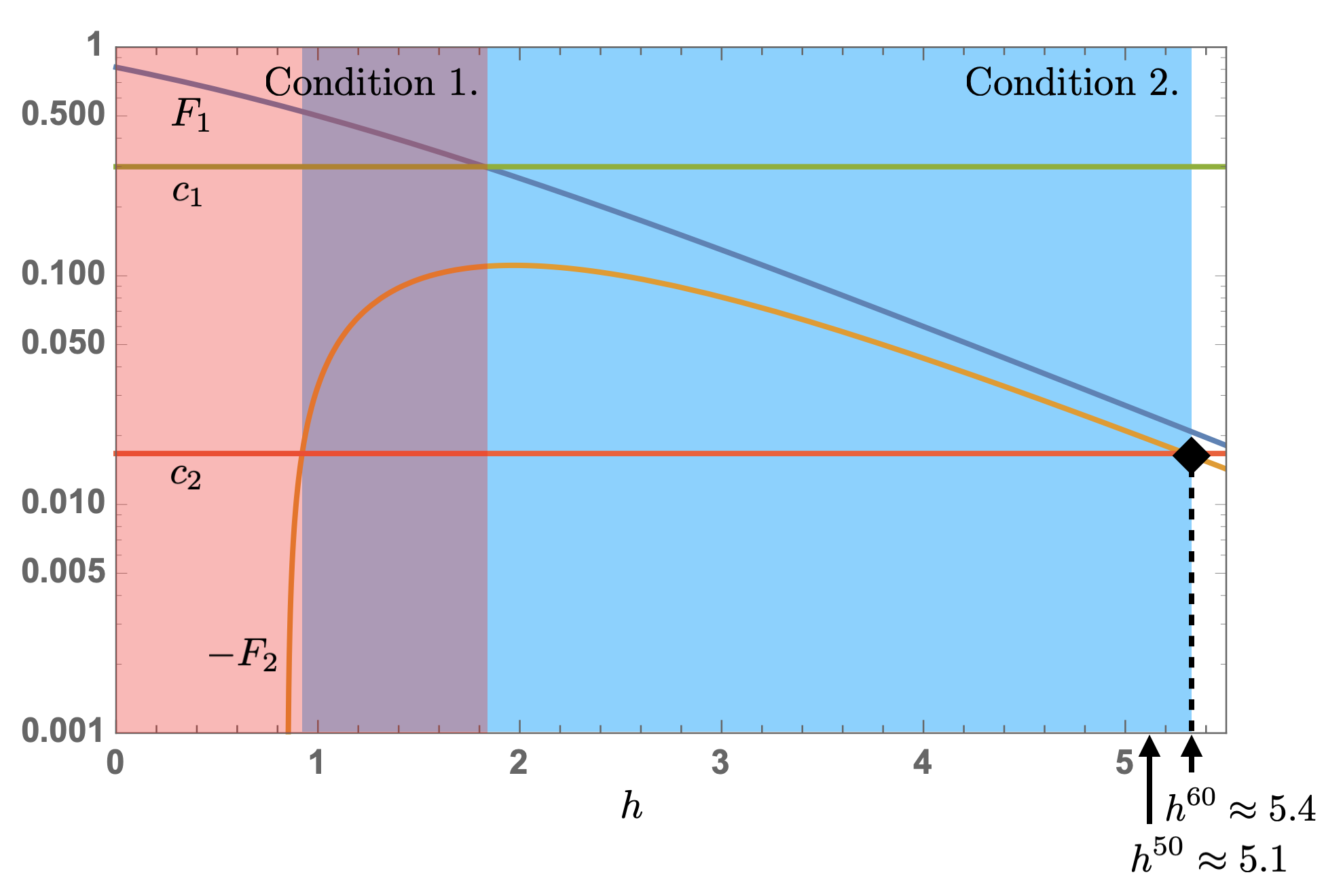}
\caption{(Case-1)(left) and (Case-2)(right) satisfying the dS conjecture below the inflation scale.}
\label{fig:cases} 
\end{figure*}
%

%=====================================================
\section{Phenomenological request for the \lowercase{d}S conjecture}
\label{sec:pheno}
%=====================================================

To identify the highest field value for $h$, we first check the number of e-folds, $N_e(h)$ and find $h_*$ giving $N_e=60 (50)$.
Assuming $h_{\rm end}\ll h_*$, the number of e-folds is given as
\begin{eqnarray}
N_{e}(h_*) 
	%&=\int_{h_{\rm end}}^{h_*} \frac{V}{dV/dh_J}\left(\frac{dh}{dh_J}\right)^2 dh_J \nonumber \\
	&\approx \frac{3}{4} e^{\sqrt{2/3}h_*} + \frac{\sqrt{6}}{4}h_*,    \label{eqn:N}
\end{eqnarray}
and for 60(50) e-folds, we get $ h_* \approx 5.4(5.1)$, respectively~\cite{Rubio:2018ogq}.
Note that this result is independent of the numerical choices of  $\lambda$ and $\xi$. 
	
By restricting our field space to $h \in \left[0, h_{*} \right]$, there can be two cases in which the Higgs potential can satisfy the conditions:

\begin{itemize}
	\item[] (Case-1) $ \phi^{*} \geq h_* \Leftrightarrow c_{1} \leq F_{1}(h_*) $
	\item[] (Case-2) $ \phi^{\star}_{\rm max} \geq h_*$ and  $\phi^{*} \geq \phi^{\star}_{\rm min} \\ ~~~~~~~~~~~~~~~\Leftrightarrow  c_{2} \leq -F_{2}(h_*) $ and $c_{1} \leq F_{1}(\phi^{\star}_{\rm min}).$
\end{itemize}

In (Case-1), the 1st condition is satisfied in the whole region, and the 2nd condition results in an arbitrary $ c_{2} $. In (Case-2), contrarily, the plateau region of the potential satisfies the 2nd condition and the low energy regions of the potential fulfill the 1st condition. 
In Fig.~\ref{fig:cases}, we schematically depict the (Case-1) and (Case-2) when the dS conjecture is fulfilled by some proper choices of $(c_1,c_2)$. Violation to these requirements will result in a region on the potential that neither satisfies both conditions, in which is contradictory to the conjecture and will result in the theory living in the Swampland. 

\begin{figure}[th]
\centering % \begin{center}/\end{center} takes some additional vertical space
\includegraphics[width=.47\textwidth]{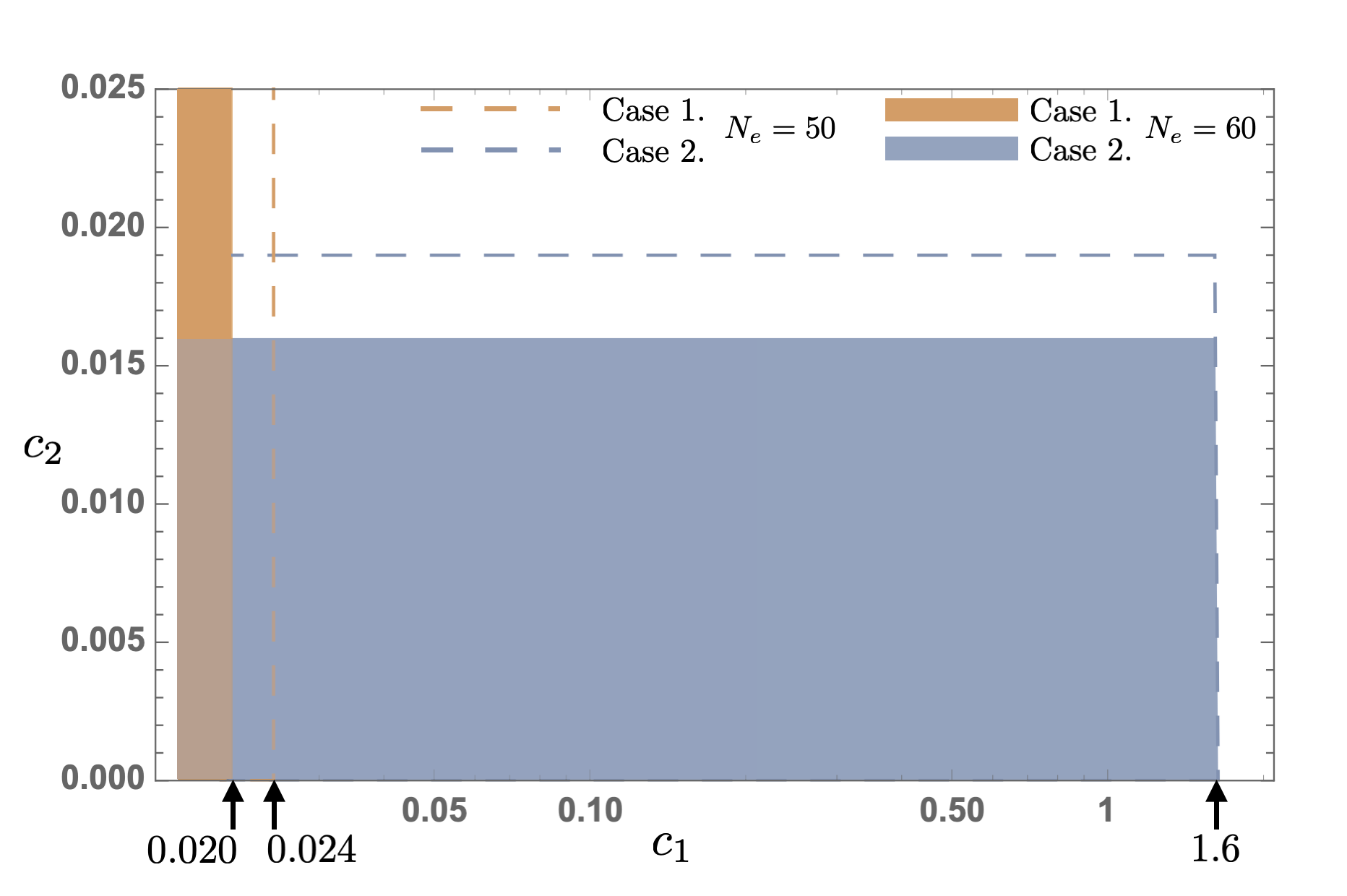}
% "\includegraphics" is very powerful; the graphicx package is already loaded
\caption{The critical line of $\left(c_1, c_2\right)$ for the Higgs inflation potential. %for $\xi = 17,000$. 
The interior(highlighted) region satisfies the dS conjecture, making the exterior of the diagram in the Swampland.}
\label{fig:c1c2} 
\end{figure}

In Fig.~\ref{fig:c1c2}, the region satisfying the dS conjecture (i.e. the region of $(c_1,c_2)$ in the ``Landscape") is depicted. (Case-1) has a bound in $c_1\lsim 0.02$. (Case-2) shows a bound for $c_1\lsim 1.6, ~ c_2 \lsim 0.016$ for $N_e = 60$. The allowed parameter space is slightly enlarged when we take $N_e=50$ instead of $60$. The parameteric dependence on $\xi$ is weak. This implies that our analysis holds for a broad parametric region of $ \xi $.
\footnote{ This low dependence on the parameters to the inflationary dynamics and the conjecture is equivalent to the Starobinsky case~\cite{Fukuda:2018haz}} 

\begin{figure*}[tbp]
\centering % \begin{center}/\end{center} takes some additional vertical space
\includegraphics[width=.47\textwidth]{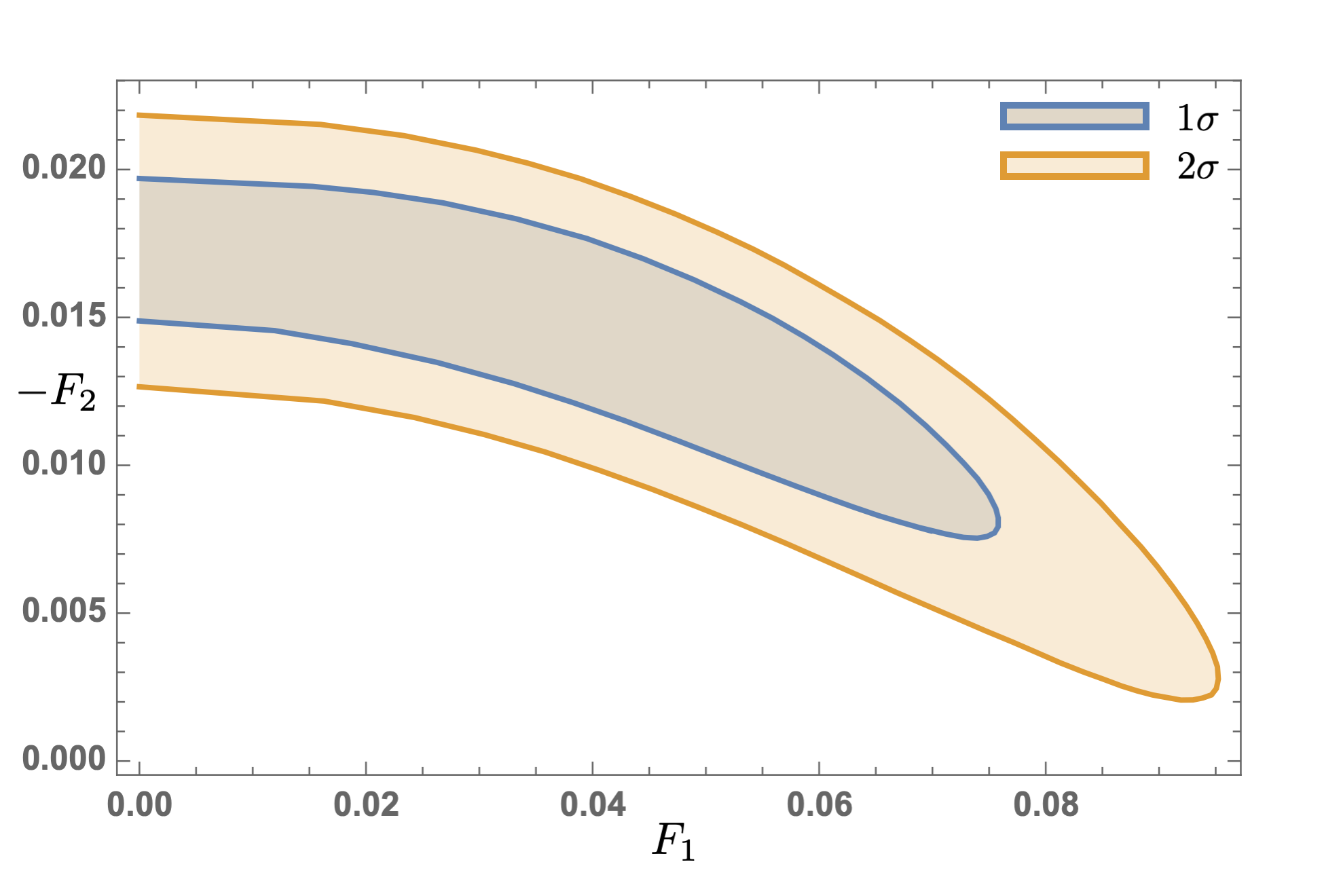}
\includegraphics[width=.47\textwidth]{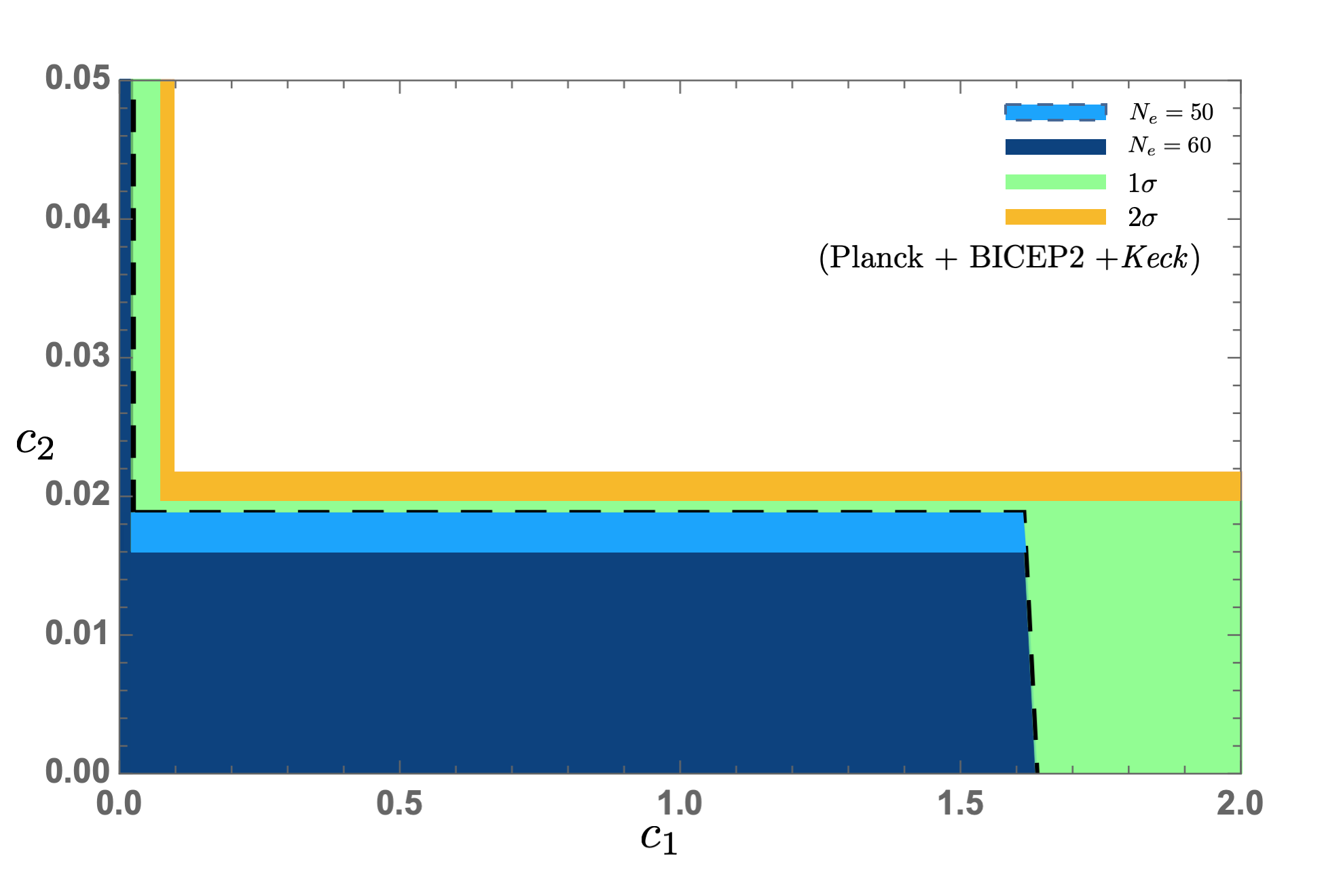}
\caption{(left)The observational bound of $\left(F_1(h_*), -F_2(h_*)\right)$ from the Planck+BICEP2+{\it Keck} observation data, from which we can read out the upper bound s on $c_1$ and $c_2$, respectively.
(right)The translated bound on  $\left(c_1, c_2\right)$ combined with the bound from the dS conjecture for the Higgs inflation model.}
\label{fig:obs} 
\end{figure*}

Having extracted the parametric regions solely from the Einstein frame version of the potential Eq.~(\ref{fig:VJ}), we turn our interest to the inflationary parameters and observational bounds given by recent observations. 
%\begin{figure}[tbp]
%\centering % \begin{center}/\end{center} takes some additional vertical space
%\includegraphics[width=.47\textwidth]{fig_higgs_c1_c2_2.png}
% "\includegraphics" is very powerful; the graphicx package is already loaded
%\caption{The critical line of $\left(c_1, c_2\right)$ for the Higgs inflation potential with $N=50$, $N=60$.}
%\label{fig:c1c2_2} 
%\end{figure}
%
%
In the slow roll region, the  tensor-to-scalar ratio and the spectral index are expressed in terms of $ (\epsilon_V, \eta_V) $ or $(F_1, F_2)$ at $h_*$ as
\begin{align}
&r = 16 \epsilon_{V} = 8F_1\left(h_*\right)^2,  \label{eqn:r}\\
&n_s = 1-6\epsilon_{V}+2\eta_{V} = 1 - 3F_1\left(h_*\right)^2 + 2F_2\left(h_*\right), \label{eqn:ns}
\end{align}
or
\begin{align}
F_1(h_*) &=\sqrt{2\epsilon_{V}}= \sqrt{\frac{r}{8}}, \\
F_2(h_*) &=\eta_{V}=\frac{n_s-1+3r/8}{2}.
\end{align}
Therefore by considering the observed results for $r$ and $n_s$, we can directly obtain the allowed region of $ (F_1(h_*),-F_2(h_*)) $ as shown in the left figure of Fig.~\ref{fig:obs}.

%
%\begin{eqnarray}
%\left.F_1(h_*)\right|_{\rm Planck+BICEP2+{\it Keck}} &\lsim& 0.09 ,\\
%\left.F_2(h_*)\right|_{\rm Planck+BICEP2+{\it Keck}} &\lsim& -0.006,
%\end{eqnarray}
%
% The upperbound of c2 is obatined from -F2(=-\eta) at r=0, not r=0.06. So above values of F1 and F2 may give some confusion. So that we have fig of \epsilon and \eta, it could provide equivalent information we want to deliver.
Here we have used recent inflationary observation data of Cosmic Microwave Background radiation (CMB) from the Planck observatory~\cite{Akrami:2018odb, Aghanim:2018eyx} and also the data taken by the BICEP2/Keck CMB polarization experiments \cite{Ade:2018gkx} 

\begin{eqnarray}
n_s \simeq 0.965 \pm 0.0004,~~~~ r\lsim0.06.
\end{eqnarray}

%
%
%\begin{figure}[tbp]
%	\centering
%	\includegraphics[width=.47\textwidth]{fig_higgs_c1_c2_obs.png}
%	\caption{The combined bound of $\left(c_1, c_2\right)$ from the observation and the Higgs inflation model.}
%	\label{fig:obs} 
%\end{figure}
%

The right figure in Fig.~\ref{fig:obs} shows the observational bound of $\left(c_1, c_2\right)$ as well as the bound of the dS conjecture for the Higgs inflation model. The observational bound is obtained by taking the maximum horizontal and vertical values from the left figure in Fig.~\ref{fig:obs}. 
The values for the observation take a contour in which  the values are 
%$\left(F_1, -F_2\right)_{1\sigma} \simeq (0.076, 0.020)$,  $\left(F_1, -F_2\right)_{2\sigma} \simeq (0.095, 0.022)$ 
$ F_{1}(h_*)\simeq0.076(0.095), -F_{2}(h_*)\simeq0.020(0.022)$ from the $ 1\sigma(2\sigma)$ confidence level of the data, and the permitted region of $c_1$($c_2$) is the highlighted area left(below) of these values.
This observational bound applies for any slow-roll inflationary models albeit the observational bound is weak in that it does not provide the upper bound of $ c_{1} $, $ c_{2} $, whereas the bound from the Higgs inflation model does for $ c_{1} $. Even when we only take the observational bound, $ c_{1} $ and $ c_{2} $ cannot coexist in the region of $ \mathcal{O}(1) $, implying the allowed region is not typically in favor with existing quantum gravity models. Meanwhile, if we take the conjecture `and' the observational bounds into account, the Higgs inflation is necessarily in the Landscape.

\vspace{.5cm}

%aba
%=====================================================
\section{Conclusion}
\label{sec:conclusion}
%=====================================================

In this letter we closely examined the Higgs inflation with nonminimal coupling in the de Sitter derivative conjecture. We evaluated the universal constants $c_1$ and $c_2$ for the Higgs potential and also provided a parametric region in which this inflation theory is to be in the Landscape. We also presented an observational bound from recent data, and the bound from the Higgs inflation model is in the interior of the observational bound. The observationally consistent values of $ c_{1} $ and $ c_{2} $ cannot simultaneously be $\mathcal{O}(1)$. However, emphasizing them being unknown parameters we can see that the Higgs inflation can be still in the Landscape. \\
%\vspace{1.0cm}

%======================================================================
%{\bf Acknowledgements:}
\acknowledgments
SCP would thank Kohei Kamada and Ryusuke Jinno for useful comments and Matt Reece for discussion during the CERN-TH workshop ``Physics at the LHC and beyond" in summer, 2018. This work was supported by the National Research Foundation of Korea (NRF) grant funded by the Korean government (MSIP) (No. 2016R1A2B2016112) and (NRF-2018R1A4A1025334).
The work of SML was supported in part by the Hyundai Motor Chung Mong-Koo Foundation.
\vspace{1.0cm}

\bibliography{hsbib_new}
\bibliographystyle{apsrev4-1}
%\nocite{*}

\end{document}